\documentclass[a4paper,10pt]{article}

\usepackage{style/osameet3}

\usepackage{amsmath,amssymb}
\usepackage[colorlinks=true,bookmarks=false,citecolor=blue,urlcolor=blue]{hyperref}
\usepackage[capitalize]{cleveref}
\usepackage[acronym]{glossaries}
\usepackage{caption}
\usepackage{subcaption}
\usepackage{xspace}
\DeclareMathAlphabet{\mathcal}{OMS}{cmsy}{m}{n}
\DeclareMathAlphabet{\mathbfit}{T1}{ptm}{b}{it}

\newcommand{\Ccal}{\mathcal{C}}
\newcommand{\Rbb}{\mathbb{R}}

\newcommand{\DOF}{\abb{DOF}\xspace}
\newcommand{\DOFs}{\abbpl{DOF}\xspace}

\newcommand{\abb}[1]{\gls*{#1}} 
\newcommand{\Abb}[1]{\Gls*{#1}} 
\newcommand{\abbpl}[1]{\glspl*{#1}} 

\newacronym{PS}{PS}{probabilistic shaping}
\newacronym{GS}{GS}{geometric shaping}
\newacronym{AIR}{AIR}{achievable information rate}
\newacronym{SSMF}{SSMF}{standard single-mode fiber}
\newacronym{SSFM}{SSFM}{split-step Fourier method}
\newacronym{EDFA}{EDFA}{erbium-doped fiber amplifier}
\newacronym{AWGN}{AWGN}{additive white Gaussian noise}
\newacronym{WDM}{WDM}{wavelength division multiplexing}
\newacronym{PAS}{PAS}{probabilistic ampltitude shaping}
\newacronym{MB}{MB}{Maxwell-Boltzman}
\newacronym[firstplural={degrees of freedom (DOFs)}]{DOF}{DOF}{degree of freedom}
\newacronym{BER}{BER}{bit error rate}
\newacronym{FEC}{FEC}{forward error correction}
\newacronym{GMI}{GMI}{generalized mutual information}
\newacronym{MI}{MI}{mutual information}
\newacronym{NLIN}{NLIN}{nonlinear interference noise}
\newacronym{PM}{PM}{polarization multiplexed}
\newacronym{QAM}{QAM}{quadrature amplitude modulation}
\newacronym{ASK}{ASK}{amplitude-shift keying}
\newacronym{PCS}{PCS}{probabilistic constellation shaping}
\newacronym{SE}{SE}{spectral efficiency}
\newacronym{GSS}{GSS}{geometric shell shaping}
\newacronym{PAPR}{PAPR}{peak-to-average power ratio}
\newacronym{BRGC}{BRGC}{binary reflected Gray code}
\newacronym{BICM}{BICM}{bit-interleaved coded modulation}
\newacronym{DSP}{DSP}{digital signal processing}
\newacronym{4D}{4D}{four-dimensional}
\newacronym{OSNR}{OSNR}{optical signal-to-noise ratio}
\newacronym{SNR}{SNR}{signal-to-noise ratio}
\newacronym{LLR}{LLR}{log-likelihood ratio}
\newacronym{BMD}{BMD}{bit-metric decoding}
\newacronym{HD}{HD}{hard-decision}
\newacronym{SD}{SD}{soft-decision}
\newacronym{SCC}{SCC}{staircase code}
\newacronym{BCH}{BCH}{Bose-Chaudhuri-Hocquenghem}

\begin{document}

\title{4D Geometric Shell Shaping with\\Applications to 400ZR}
\vspace{-2ex}
\author{Sebastiaan Goossens$^{1}$, Yunus Can Gültekin$^{1}$, Olga Vassilieva$^{2}$, Inwoong Kim$^{2}$, \\ Paparao Palacharla$^{2}$, Chigo Okonkwo$^{1}$, Alex Alvarado$^{1}$}
\address{$^{1}$Department of Electrical Eng., Eindhoven University of Technology, 5600MB, Eindhoven, The Netherlands\\
$^{2}$Fujitsu Network Communications, Inc., 2801 Telecom Pkwy., Richardson, Texas, USA}\vspace*{-0.5ex}
\email{s.a.r.goossens@tue.nl}

\vspace*{-3.3ex}
\begin{abstract}
Geometric shell shaping is introduced and evaluated for reach increase and nonlinearity tolerance in terms of MI against PM-16QAM and PS-PM-16QAM in a 400ZR compatible transmission setup.
\end{abstract}

\vspace*{0ex}
\section{Introduction}
\vspace*{-1.5ex}
In recent years, \abb{GS} has been widely investigated to cope with the exponentially growing demand for optical fiber capacity.
Conventionally, geometrically shaped constellations are designed in the 2D (complex) plane under the \abb{AWGN} channel assumption\cite{Qu2019}. 
However, it has been shown that multidimensional constellation design is able to better reduce \abb{NLIN} with respect to 2D constellations in addition to providing shaping gain \cite{Kojima2017}. In this paper, a novel framework called 4D \abb{GSS} is introduced for designing well-structured constellations which overcome the complexities of the unconstrained 4D constellation optimization for the nonlinear fiber channel. We then design constellations in the 4D GSS framework for a 400ZR compatible channel model with high \abb{NLIN}.

\vspace*{-2ex}
\section{4D Geometric Shell Shaping}
\vspace*{-1.25ex}
We consider a 4D constellation $\mathcal{C}=\{\mathbfit{x}_i:i=1,\ldots,M\}$ where $\mathbfit{x}_i\triangleq (x_{1i},x_{2i},x_{3i},x_{4i})\in\mathbb{R}^4$. Here, $M=2^m=|\mathcal{C}|$, and $x_{1i},x_{2i}$ and $x_{3i},x_{4i}$ represent the 2D coordinates of the x- and y-polarization, respectively. Polarization multiplexed 16-ary quadrature amplitude modulation (PM-16QAM) is an instance of this where $m=8$ and $x_{1i},x_{2i},x_{3i},x_{4i}\in \{\pm1,\pm3\}$. Each point in $\mathcal{C}$ has four \DOFs. Geometrically optimizing a 4D constellation in an unconstrained way results in $4\cdot2^m$ \DOFs, which quickly becomes challenging as $m$ increases. To introduce structure in the constellation, and hence, to reduce the number of \DOFs, we propose to impose three constraints: The constellation has (i) orthant symmetry, and (ii) X-Y symmetry, and (iii) the constellation points are uniformly divided on a limited number of 4D shells. We call the optimization under these constraints  4D-\abb{GSS}.

An orthant is a generalization in $N$D Euclidean space of what a quadrant is in the 2D plane, with $2^N$ orthants in total. The first orthant is the region $\Rbb^N_+$ where each dimension has a positive value. (i) A constellation $\Ccal$ is \textit{orthant symmetric} if every point is mirrored across all $N$ dimensions. The points in the first orthant are defined as $\Ccal_+ = \{\mathbfit{x}_j:j=1,\ldots,2^{m-N}\} \in \Rbb_+^N$. This reduces the amount of \DOFs by a factor of $2^N$. In the context of optical communications, orthant symmetry was used and explained in \cite[Sec. II-B]{Chen9345386}. (ii) $\mathcal{C}$ is \textit{X-Y symmetric} if whenever $(x_{1i},x_{2i},x_{3i},x_{4i}) \in \mathcal{C}$, then so is $(x_{3i},x_{4i},x_{1i},x_{2i}) \in \mathcal{C}$. This reduces the number of \DOFs by a factor of 2.
(iii) A constellation $\Ccal$ is said to be {\it uniformly divided k-shell constrained} if $||\mathbfit{x}_i||\in \{r_1,\ldots,r_k$\} where $k=2^p$ is the number of shells, with each shell having the same amount of points, $i=1,\ldots,M$, and $p\in\mathbb{N}$. By forcing each point to be on a certain shell, we take away one additional \DOF per constellation point, such that $N-1$ \DOFs per constellation point remain. However, $k$ extra \DOFs are added due to the number of shells. The total amount of \DOFs is now $(N-1)\cdot2^{m}+k$ (instead of $N\cdot2^m$) by applying these three constraints.

The constraints above are applied to a 4D constellation targeting the same spectral efficiency as PM-16QAM ($m=8$). After applying the first two constraints, the amount of \DOFs is reduced by a factor of $2^5$. This constellation is denoted by $\Ccal_+'$. Shell constraints are applied to $\Ccal'_+$ using $k=4$ shells, which results in the constellation called 4D-GSS-4. The remaining 8 points in this constellation can be expressed in $3\cdot2^{m-5}+k = 28$ \DOFs.

\vspace*{-2ex}
\section{System Setup and Optimization}
\vspace*{-1.25ex}
We consider an unamplified 400ZR \cite{400GZR} system which results in a high-\abb{NLIN} scenario. A dual-polarized, single channel waveform is transmitted over a single span of standard single mode fiber (SSMF). We simulate this setup via the split-step Fourier method (SSFM) for which the symbol rate is matched to the 400ZR specification at 59.84 GBd. Since an unamplified link is used, noise sources due to transmitter and receiver impairments are emulated instead of simulating optical amplification. For this, worst-case design parameters from the 400ZR specification are used as a guideline. At the transmitter side, the 400ZR specification requires the in-band \abb{OSNR} to have a minimum value of 34 dB/0.1nm. At the receiver side, the concatenated \abb{FEC} scheme in the 400ZR standard is specified to operate error-free (post-\abb{FEC} \abb{BER} = $10^{-15}$) when a pre-\abb{FEC} \abb{BER} of $1.25\cdot10^{-2}$ or lower is achieved. The receiver sensitivity requires least $-20$ dBm of power to be present at the input of the receiver. These previous two conditions combined with the minimum of 34 dB \abb{OSNR} at the transmitter guarantee error-free operation.

\begin{figure}
    \centering
    \includegraphics{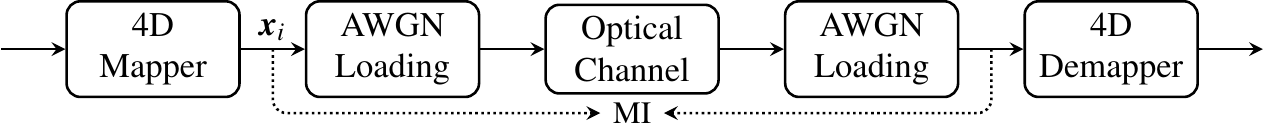}
    \vspace*{-6.1ex}
    \caption{System overview including the AWGN loading stages\vspace*{-5.9ex}}
    \label{fig:system_setup}
\end{figure}

In simulations, the transmitter is emulated by adding \abb{AWGN} to the transmitted waveform such that the \abb{OSNR} value is equal to this in-band \abb{OSNR} limit. For the receiver, it is possible to calculate the necessary \abb{AWGN} addition using \cite[eq. (18)]{752121}, which for 16QAM and a \abb{BER} of $1.25\cdot10^{-2}$ provides a \abb{SNR} of 13.5 dB. Assuming that an input power of $-20$ dBm at the receiver is obtained in a back-to-back scenario, the amount of noise power added in the receiver is equal to $-33.5$ dBm. This is added as \abb{AWGN} after simulating the optical fiber. The full system can be seen in \cref{fig:system_setup}.

To find the optimal constellation $\Ccal^*$ under the given constraints, each point in $\Ccal'_+$ is now represented in spherical coordinates $(r_i,\theta_j,\phi_j,\omega_j)$, where $i=1,\ldots,k$ and $j=1,\ldots,2^{m-5}$. The optimization problem can now be defined as $ \{\mathbfit{r}^*,\mathbfit{\theta}^*,\mathbfit{\phi}^*,\mathbfit{\omega}^*\} = \text{argmax}~\text{MI}(\mathbfit{r},\mathbfit{\theta},\mathbfit{\phi},\mathbfit{\omega},P)$, with $\mathbfit{r}:0\leq r_i\leq1$ and $\mathbfit{\theta},\mathbfit{\phi},\mathbfit{\omega}:0\leq{\theta_j},{\phi_j},{\omega_j}\leq\pi/2$, so that the optimization parameters are constrained such that $\Ccal_+'\in\Rbb^4_+$. \Abb{MI} \cite{Alvarado8240991} is chosen as the performance metric and the launch power is denoted by $P$. 
The parameters are initialized to be halfway between the bounds, and optimization is performed over the proposed system using a patternsearch optimizer \cite{patternsearch}, which is a derivative-free multidimensional optimization algorithm. 

\vspace*{-2ex}
\section{Results}
\vspace*{-1.25ex}
Conventional PM-16QAM, as used in the 400ZR standard, is considered as the baseline. Next to that, \abb{PS} is applied on top of PM-16QAM by shaping each real dimension with an ideal amplitude shaper, where amplitudes are randomly drawn from a predefined distribution. The distribution is optimized for each $P$ and transmission distance combination and is not constrained to be a Maxwell-Boltzmann distribution.
Lastly, an AWGN optimal 4D sphere packed constellation (w4-256) \cite{w4_256} is simulated.

\begin{figure}[h]
    \begin{subfigure}[b]{0.55\textwidth}
        \centering
        \includegraphics{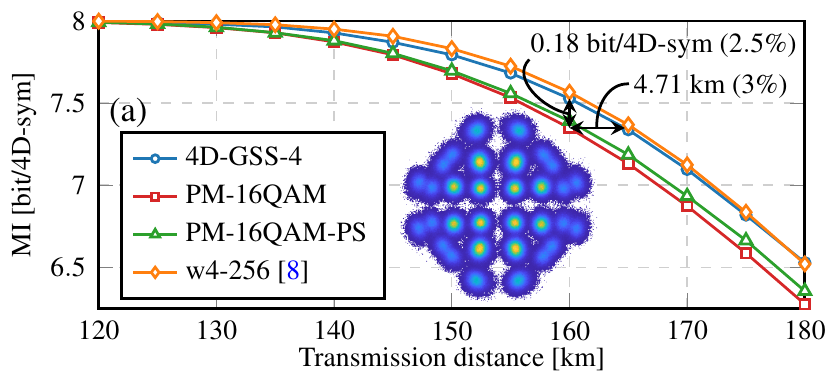}
        \caption{}
        \label{fig:MI_dist}
    \end{subfigure}%
    \begin{subfigure}[b]{0.45\textwidth}
        \centering
        \includegraphics{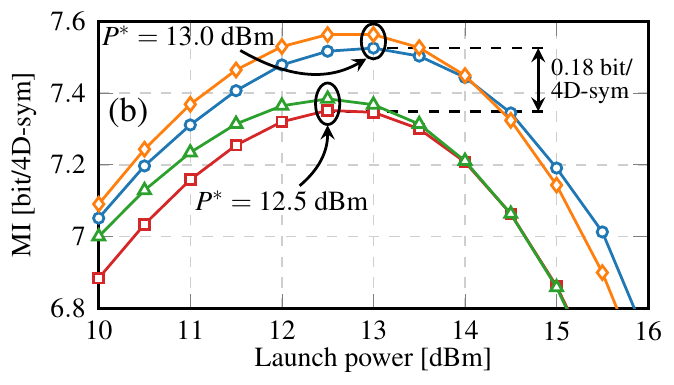}
        \caption{}
        \label{fig:MI_pow}
    \end{subfigure}%
    \vspace{-8ex}
    \caption{MI vs distance at optimal launch power (left) and launch power at $160$~km (right). Constellation inset shows a 2D projection of the received symbols at the optimal launch power for a distance of $120$~km.\vspace*{-3ex}}
    \label{fig:results}
\end{figure}

\Cref{fig:MI_dist} shows \abb{MI} results for a distance sweep where 4D-GSS-4 has $3\%$ gain in distance and $2.5\%$ gain in MI vs PM-16QAM around 160 km and is slightly outperformed by w4-256. 
\Cref{fig:MI_pow} shows that 4D-GSS-4 has the largest optimal launch power $P^*$ and hence, the highest nonlinear tolerance among the considered schemes. Moreover, due to the rapidly-vanishing MI of w4-256, 4D-GSS-4 outperforms w4-256 at very high powers ($P>14$). However, for lower powers ($P<14$), w4-256 achieves larger MI than 4D-GSS-4. Finally, we have numerically confirmed that 4D-GSS-4 achieves a $49\%$ lower \abb{PAPR} ratio on average vs PM-16QAM and $10\%$ lower \abb{PAPR} vs w4-256.

Even though w4-256 generally performs the best in terms of of MI, we conjecture that the orthant symmetry in 4D-GSS-4 allows for very good (Gray) labeling, resulting in high GMI performance. Moreover, the $k$ number of shells allow for straightforward implementation of \abb{PS} on the 4D shells, enabling rate-adaptation.

\vspace*{-2.5ex}

\vspace*{-3.77ex}
\section{Conclusions}
\vspace*{-1.25ex}
A novel framework is proposed for generating families of well-structured 4D geometrically-shaped constellations which are more nonlinearity-tolerant than conventional PM-16QAM. The newly proposed 4D-GSS-4 constellations outperform both PM-16QAM and PS-PM-16QAM in a 400ZR compatible transmission setup.\\
\vspace*{-2.25ex}

\footnotesize{
\noindent\textbf{Acknowledgements:}~The work of Sebastiaan Goossens, Yunus Can Gültekin and Alex Alvarado has received funding from the European Research Council under the European Union's Horizon 2020 research and innovation programme via the Starting grant FUN-NOTCH (grant agreement ID: 757791) and via the Proof of Concept grant SHY-FEC (grant agreement ID: 963945).}

\end{document}